# Nine Million Book Items and Eleven Million Citations: A Study of Book-Based Scholarly Communication Using OpenCitations

Yongjun Zhu[1], Erjia Yan[2], Silvio Peroni[3], and Chao Che[4*]

1. Department of Library and Information Science, Sungkyunkwan University, 25-2, Sungkyunkwan-ro, Jongno-gu, Seoul, Republic of Korea. yzhu@skku.edu

2. College of Computing and Informatics, Drexel University, 3141 Chestnut Street, Philadelphia, PA 19104, USA. ey86@drexel.edu

3. Digital Humanities Advanced Research Centre (DHARC), Department of Classical Philology and Italian Studies, University of Bologna, Via Zamboni 32, 40126 Bologna, Italy. silvio.peroni@unibo.it

4. Key Laboratory of Advanced Design and Intelligent Computing, Ministry of Education, Dalian University, Dalian, China. chechao101@163.com

*Corresponding author

**Abstract**
Books have been widely used to share information and contribute to human knowledge. However, the quantitative use of books as a method of scholarly communication is relatively unexamined compared to journal articles and conference papers. This study uses the COCI dataset (a comprehensive open citation dataset provided by OpenCitations) to explore books' roles in scholarly communication. The COCI data we analyzed includes 445,826,118 citations from 46,534,705 bibliographic entities. By analyzing such a large amount of data, we provide a thorough, multifaceted understanding of books. Among the investigated factors are 1) temporal changes to book citations; 2) book citation distributions; 3) years to citation peak; 4) citation half-life; and 5) characteristics of the most-cited books. Results show that books have received less than 4% of total citations, and have been cited mainly by journal articles. Moreover, 97.96% of books have been cited fewer than ten times. Books take longer than other bibliographic materials to reach peak citation levels, yet are cited for the same duration as journal articles. Most-cited books tend to cover general (yet essential) topics, theories, and technological concepts in mathematics and statistics.

**Introduction**

Books are one of the most traditional writing formats. As a genre, they have been used to deliver mass academic knowledge on a variety of subjects in a well-organized way. Books deliver diverse contents at different levels, ranging from fundamental knowledge underpinning studies, to collections of research articles on a specific topic. Therefore, books have been widely used by scientists and scholars to share their discoveries. Through this process, books contribute to the accumulation of scientific knowledge (Serenko, Bontis, & Moshonsky, 2012).

Despite their long history as a medium for information delivery, books' roles in scholarly communication are relatively unexamined compared to their journal counterparts. Citations are one of the most useful tools for measuring the reach of scientific publications and analyzing scholarly communication. Thus, the comparative lack of book citation data is a significant reason for our limited understanding of their usage in academic discourse (Adam, 2002; Garfield, 1972; Moed, 2005). Recently, with the emergence of a few bibliographic databases and services that support book citation data, they have been studied in a research evaluation context. The motivation for these studies stems from the notion that books, along with other types of publications, are an essential measure of individuals' research achievements, in particular in the social science and humanities fields (Bott & Hargens, 1991; Huang & Chang, 2008; Nederhof, 2006). Accordingly, researchers have studied different ways to assess books' impact using citation data (e.g., Kousha & Thelwall, 2018; Zuccala, Breum, Bruun, & Wunsch, 2018). Data from Google Books (Kousha & Thelwall, 2009), Google Scholar (Kousha, Thelwall, & Rezaie, 2011), the Web of Science Book Citation Index (Leydesdorff & Felt, 2012a; Torres-Salinas, Robinson-García, Cabezas-Clavijo, & Jiménez-Contreras, 2014; Zuccala et al., 2018), Scopus (Kousha et al., 2011), and Microsoft Academic (Kousha & Thelwall, 2018) has been used to assess book citation impact.

Since the goal of the abovementioned studies is to evaluate research, the standard unit of analysis has been individual books. Therefore, despite the recent studies of book citations, the roles of books as a medium for scholarly exchange remains an open question. Understanding the roles of books in scholarly communication is an important and timely question. Highly-specialized research fields, easy accessibility of research resources, and short research publication cycles characterize modern science. Compared to other publications such as journal articles and conference papers, books tend to have broader focuses and more general audiences (Glänzel, Thijs, & Chi, 2016). A book may serve as a fundamental knowledge source for a wide range of research studies. Advances in academic databases and search engines have greatly improved online access to research resources (Zhu, Yan, & Song, 2017). Although books are more easily accessible than ever before, print is still their main publication format. As yet, they have not taken full advantage of modern information systems. The publication cycle of books is usually relatively long. Both authors and publishers need substantial time to plan, write, edit and publish a book. Books' in-depth coverage of a subject also prolongs their publication cycle. The long publication cycle negatively affects the value of books dealing with time-sensitive topics. By the time of publication, their contents may be outdated. The differing characteristics of books versus journal articles and conference papers may mean that each plays a different role in knowledge propagation. More investigation is needed to expand our knowledge and understanding of books' purpose in academic dissemination.

A comprehensive citation dataset (including books and other types of research publications) is required in order to understand the role of books in scholarly communication. By analyzing this data, we can compare the roles of books with those of other types of research publications. Most existing citation datasets concern one type of research publication. Consequently, the lack of this kind of comprehensive citation dataset had been a significant barrier to our understanding of books' roles. Besides, most current citation analyses use data from proprietary sources such as Clarivate's Web of Science (WoS) and Elsevier's Scopus. Use of these closed data sources has made it quite challenging to share materials, thus hindering peer researchers' abilities to reproduce, replicate, or build upon existing investigations. Citation analysis is highly dependent on large-scale bibliographic data. Hence, the use of open data is a pressing need. Open data does more than enable researchers free access to, and usage of, information. It also facilitates a paradigm shift in modes of inquiry by advancing open and transparent dialogue (as prescribed by shared open science practices). Various stakeholders, including researchers, evaluators, and science policymakers, stand to benefit from such availability.

OpenCitations (http://opencitations.net) is a small, independent scholarly infrastructure organization (Peroni & Shotton 2019). It is dedicated to open scholarship, as well as the publication of open bibliographic and citation data using Semantic Web (Linked Data) technologies. It also engages in advocacy for semantic publishing. Moreover, it is a crucial member of the Initiative for Open Citations (I4OC, https://i4oc.org). It provides, maintains and updates the OpenCitations Data Model (Peroni and Shotton, 2018a), which is based on SPAR (Semantic Publishing and Referencing) Ontologies (Peroni and Shotton, 2018b) (http://www.sparontologies.net). SPAR may be used to encode all aspects of scholarly, bibliographic, and citation data in the Resource Description Framework (RDF), enabling them to be published as Linked Open Data (LOD). Separately, OpenCitations provides open-source software with generic applicability for searching and browsing. This software also provides application programming interfaces (APIs) over the RDF triplestores (https://github.com/opencitations). It has developed the OpenCitations Corpus (OCC, http://opencitations.net/corpus) (Peroni, Shotton, & Vitali, 2017), a database of open downloadable bibliographic and citation data. Said data was recorded in RDF and released under a Creative Commons (CC0) public domain waiver, and the OpenCitations Index of Crossref open DOI-to-DOI citations (COCI) (http://opencitations.net/index/coci) (Heibi, Peroni, & Shotton, 2019). COCI presently contains information on more than 445 million citations, released under a CC0 waiver.

Using the COCI data made available by OpenCitations, the Crossref metadata accessible via their API (https://api.crossref.org), and the categories in the Library of Congress Classification retrievable using the OCLC API (http://classify.oclc.org/classify2/api_docs/index.html), this study aims to understand the roles of books in scholarly communication, employing the following research questions: First, how have citations to books changed over time? Second, how citation patterns of books are different from that of journal articles and conference papers? Finally, what are the characteristics of most-cited books?

**Data and Methods**

We used the latest COCI data dump (November 2018) (OpenCitations, 2018), which includes more than 445 million citation links between more than 46 million bibliographic resources. We also used the Crossref API (https://api.crossref.org) to download all of these bibliographic resources' metadata (i.e., title, DOI, number of authors, and type of bibliographic resource). Finally, we used the OCLC API (http://classify.oclc.org/classify2/api_docs/index.html) to retrieve the Library of Congress Classification (LCC) categories of all the books included in our dataset by matching their ISBNs. The COCI data citation counts and the Crossref metadata used in this analysis have been compiled into a single CSV file, while the LCC categories associated to each book and the related mapping between such categories with the five Web of Science research areas (i.e. *Arts & Humanities*, *Life Sciences & Biomedicine*, *Physical Sciences*, *Social Sciences*, and *Technology*) have been compiled in other two distinct CSV files. All these files have been made available on Zenodo (Zhu, Y., Yan, E., Peroni, S., & Che, C, 2019). In our study, we have considered three specific types for each of the bibliographic resource included in the dataset, i.e. *book*, *journal*, and *conference*. We have assigned these three types according to the particular Crossref content type to each resource.

In particular, we classified as "book" all the entities that had one of the following Crossref type: monograph, book section, book track, book part, book set, book chapter, reference book, book series, edited book, and book. Similarly, we classified as "conference" all the entities having one of the following Crossref types: proceedings article, proceedings series, and proceedings. We considered proceedings articles as conference papers. Finally, we classified as "journal" all the entities having one of the following Crossref types: journal, journal volume, journal issue, and journal article. We included only journal articles in the study. We did not differentiate types of journal articles (article, review, and others). Table 1 shows descriptive statistics on the dataset. Because numerous bibliographic entities have never been cited, the median citation was computed using entities that have been cited at least once.

Table 1. Data Statistics

|  | Book items | Journal articles | Conference papers |
|---|---|---|---|
| Entity Size (total) | 9,307,705 | 71,170,924 | 4,541,806 |
| Entity Share (total) | 11% | 84% | 5% |
| Entity Size (no citation) | 7,698,881 | 38,340,154 | 3,032,178 |
| Entity Share (no citation) | 16% | 78% | 6% |
| Citation Size | 11,327,826 | 404,373,547 | 8,184,719 |
| Average Citation | 1.22 | 5.68 | 1.80 |
| Median Citation | 2 | 5 | 2 |
| Max Citation | 12,513 | 49,282 | 49,282 |

The sizes of the COCI CSV dump and the bibliographic resource metadata file are 76G and 13G respectively, while the size of the LCC categories associated with each book is around 26 MB. After downloading the raw data, we performed preprocessing. We removed bibliographic entities that have a negative number (e.g., -59011401600000), or a two- or three-digit number as the publication year – these mistakes derived from the Crossref metadata. We also removed citation records in which the citing bibliographic entity has a publication year earlier than the cited one. Such errors could be due to either a mistake or to a cited entity's postponed publication date, e.g. when the preprint of a bibliographic entity is cited before its formal publication in a book or journal. We also removed multiple citations to different editions of a book when such citations pointed to a later-published edition of a book. In total, we removed 3% of the total bibliographic entities and 1.4% of the total citations. We used a MongoDB instance (a NoSQL database for large volumes of data) to store all the preprocessed.

Using the data, we investigated the following aspects of book use in scholarly communication:

*Temporal Changes of Citations to Books*: We analyzed yearly citations to books to identify overall trends, as well as temporal changes. Specifically, we analyzed temporal changes of book citations compared to citations of journal articles and conference papers. We also aimed to investigate how each bibliographic type (i.e., books, journal articles, and conference papers) tends to cite books.

*Citation Distribution and the Relationship Between Publication Year and Citation*: We categorized citation counts into five ranges: 0-10, 10-50, 50-100, 100-500, and 500+. We computed the percentage of books falling into each range to discern the frequency distribution of citations. Publication year is an essential factor regarding citation counts. Thus, we analyzed the relationship between a book's publication year and the yearly average citations of books. We aimed to understand this relationship using statistical modelling.

*Years to Citation Peak and Cited Half-Life*: We computed the years to citation peak by subtracting a book's publication year from the year in which it received the most significant number of citations ($Y_{ycp} = Y_{citation\_peak} - Y_{publication}$). By measuring the years to a book's citation peak, we see how quickly it obtains its highest popularity. We also computed the book's cited half-life (Todorov & Glänzel, 1988) as the years taken to reach half of its total citations ($Y_{chl} = Y_{total\_citations/2} - Y_{publication}$). Through these two measures, we can see how long a book was cited, and how slowly or quickly citations to a book peak and dropped off.

*Disciplinary Differences in Book Citations*: we analyzed whether different disciplines, as defined by the WoS main research fields, exhibit different citation characteristics.

*The Twenty Most-Cited Books*: To summarize their characteristics, we investigated the twenty most cited books in terms of publication year, title, and citations from each bibliographic type.

## Results

*Temporal Changes of Citations to Books*

Figure 1 shows yearly publication counts of books, journal articles, and conference papers. It also indicates citations received by each of the three bibliographic types over the last fifty years.

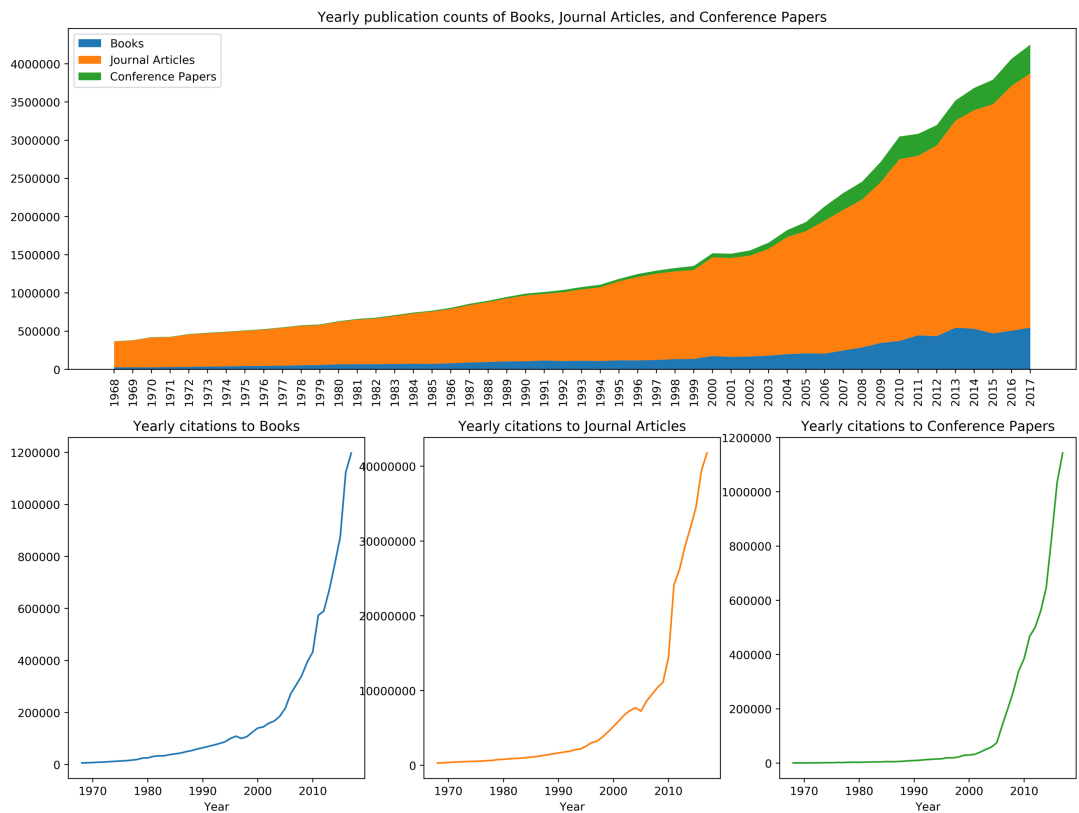

Figure 1. Yearly Publication and Citation Counts

Journal articles' yearly publication counts were much larger than the other two bibliographic types. Therefore, they were cited more frequently. After 2010, there were sharp increases in both publication and citation counts. Table 2 shows raw citations, and the percentages of each citation type between 1967 to 2017 with the step size of five.

Table 2. Citations to Books, Journal Articles, and Conference Papers (1967 – 2017)

| Year | Citations to Books (%) | Citations to Journal Articles (%) | Citations to Conference Papers (%) |
|---|---|---|---|
| 2017 | 1,281,594 (2.82%) | 42,998,567 (94.65%) | 1,149,638 (2.53%) |
| 2012 | 639,596 (2.27%) | 27,037,678 (95.94%) | 505,454 (1.79%) |
| 2007 | 333,736 (3.22%) | 9,835,239 (94.84%) | 201,854 (1.95%) |
| 2002 | 170,430 (2.36%) | 6,995,930 (97.07%) | 40,900 (0.57%) |
| 1997 | 107,703 (3.10%) | 3,351,547 (96.34%) | 19,730 (0.57%) |
| 1992 | 79,815 (3.90%) | 1,955,843 (95.48%) | 12,856 (0.63%) |
| 1987 | 52,415 (3.76%) | 1,334,051 (95.82%) | 5,852 (0.42%) |
| 1982 | 35,569 (3.61%) | 945,075 (95.97%) | 4,083 (0.41%) |
| 1977 | 17,984 (2.72%) | 642,194 (96.96%) | 2,121 (0.32%) |
| 1972 | 10,342 (2.14%) | 471,861 (97.66%) | 978 (0.20%) |
| 1967 | 6,103 (2.05%) | 291,462 (97.80%) | 441 (0.15%) |

In the past fifty years, overall citations to books constituted less than 4% of total citations. Citations to books started at 2% in the 1960s and 1970s, maintained a 3% level in the 1980s and 1990s and dropped back to 2% in the 2000s and 2010s. Among the values shown in the table, the highest was 3.90% (in 1992) while the lowest value in the past twenty years was 2.27% (2012). Citations to journal articles have decreased slightly in recent years (from 97.07% in 2002 to 94.65% in 2017). Decreased percentages of books and journal articles have been absorbed by conferences. Citations to conferences have been steadily increasing, from 0.15% in 1967 to 2.53% in 2017.

Figure 2 shows citation changes (in percentages) of each bibliographic type from 1967 to 2017.

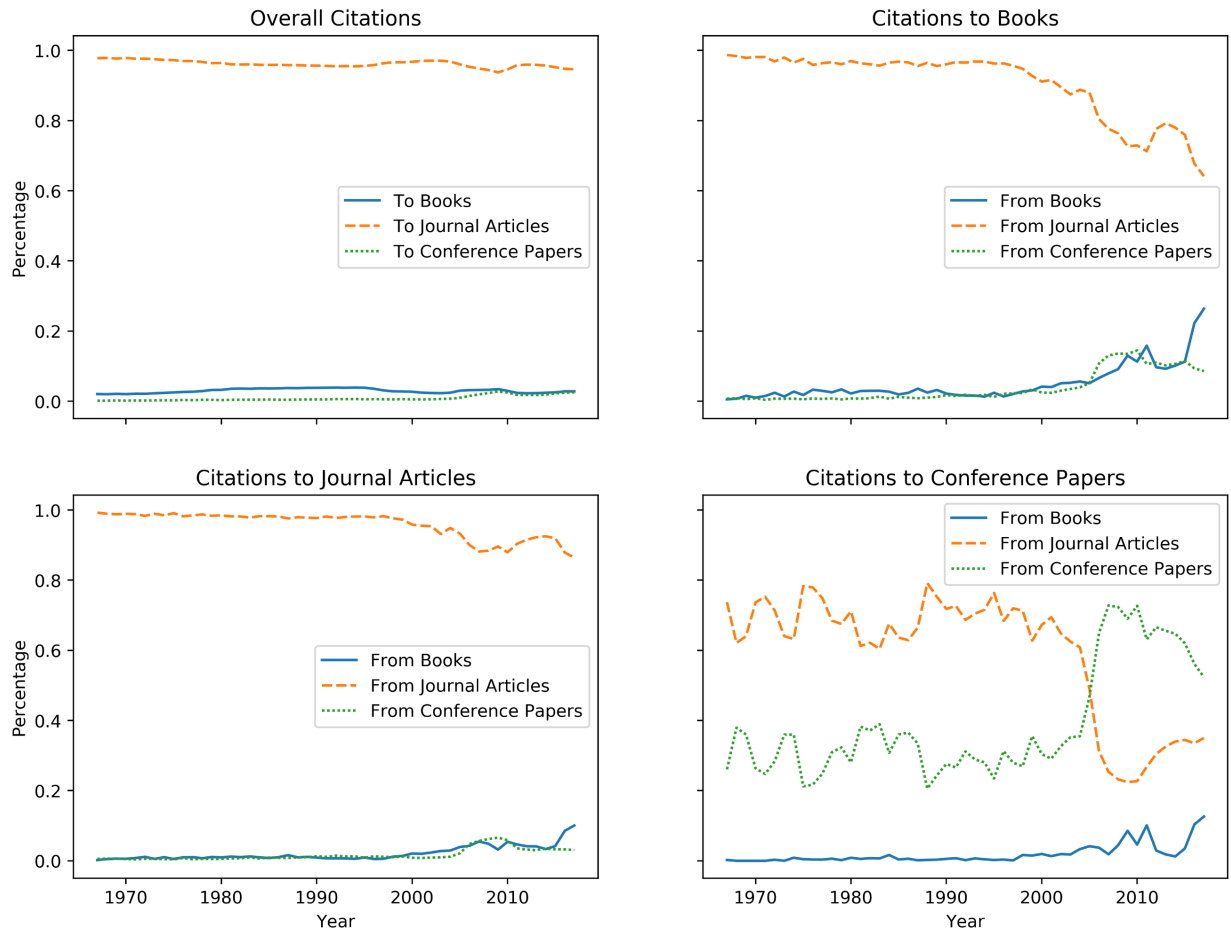

Figure 2. Citation Changes in Percentages Between 1967 and 2017

Although there have been no remarkable changes in book citations, Figure 2 shows that their percentage hit the highest point in the 1990s and continued to decrease after that (apart from slight fluctuations in the late 2000s and the early 2010s). Books received most of the citations from journal articles before the 2000s. In the 2000s and 2010s, citations from journal articles kept dropping, while books and conference papers cited books far more frequently. Specifically, citations from books grew to 20% in the late 2010s. Conference paper citations showed the same pattern, with citations from conference papers increasing, and those from journal articles decreasing. Besides, books cited conference papers more often. In the early years, journal articles were the primary source of citations for all the three bibliographic types. However, in the 2000s, conference papers replaced journal articles. In this shift, conference papers became the largest source of citations to other conference papers. Both books and conference papers cited their genre most often, whereas journal articles cite these two genres less than before.

*Citation Distribution and the Relationship Between Publication Year and Citation*

Table 3 shows the distribution of raw citation counts and percentages of five citation ranges: 0-10, 10-50, 50-100, 100-500, and 500+.

Table 3. Citation Counts and Percentages of Five Citation Ranges

| Citation Range | Books | | Journal Articles | | Conference Papers | |
|---|---|---|---|---|---|---|
| | Count | Percentage | Count | Percentage | Count | Percentage |
| 0-10 | 9,163,128 | 97.96% | 62,436,583 | 87.12% | 4,428,629 | 96.79% |
| 10-50 | 158,470 | 1.69% | 7,882,731 | 11.00% | 130,193 | 2.85% |
| 50-100 | 18,690 | 0.20% | 905,831 | 1.26% | 11,242 | 0.25% |
| 100-500 | 12,033 | 0.13% | 416,001 | 0.58% | 5,169 | 0.11% |
| 500+ | 1,385 | 0.01% | 24,953 | 0.03% | 329 | 0.00% |

All three bibliographic types in Table 3 showed a long-tailed distribution. 97.96% of books have been cited fewer than ten times, and fewer than 2% of books fall into the 10-50 citation range. The skewness in books is more significant than that in journal articles (where 87.12% received less ten citations, 11% received more than ten citations, and only a few received more than fifty citations).

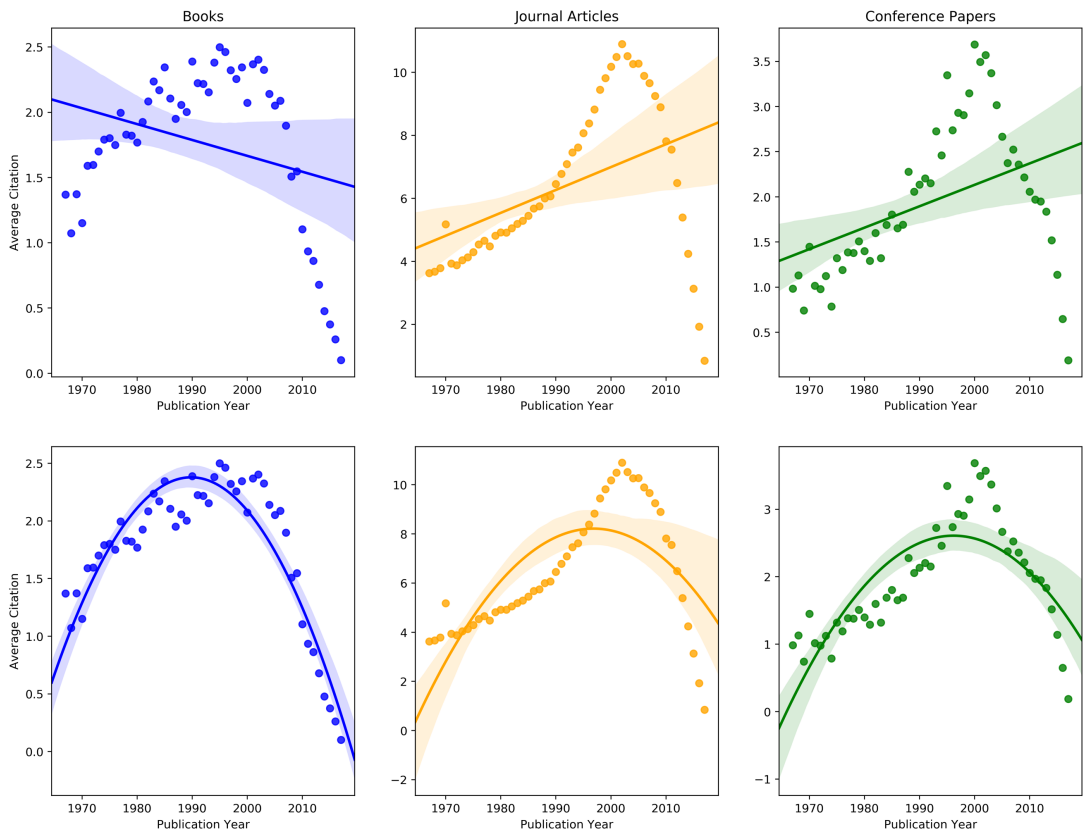

Figure 3. Relationship between Publication Year and Average Citation Count

Figure 3 shows the relationship between the publication year and the average number of citations received by each publication type. Linear (upper row) and quadratic polynomial (lower row) regression lines were fitted to model the relationship.

The pattern in books is different from those of journal articles and conference papers. In books, the linear regression line has a negative slope, while the slopes of the other two bibliographic types are positive. It is clear that, before 2000, recent journal articles and conference papers received more citations than older ones. After 2000, the opposite pattern is evident. In terms of books, the average citation is relatively stable (between 2 and 2.5) between 1980 and 2010. A close look at the patterns reveals that books published in the 1990s received more citations than books published at other times. A quadratic polynomial regression model, in which the vertex is above the 1990s, explains the pattern well.

*Years to Citation Peak and Cited Half-Life*

Publications take time to reach their citation peaks. Citation peak denotes the year in which a publication receives the highest number of citations. After its citation peak, a publication is cited fewer times, suggesting its diminishing influence. We computed years to citation peak for each book, and we show the related distributions in Figure 4. Figure 4 shows both frequency and cumulative frequency distributions of years to citation peak from one to twenty.

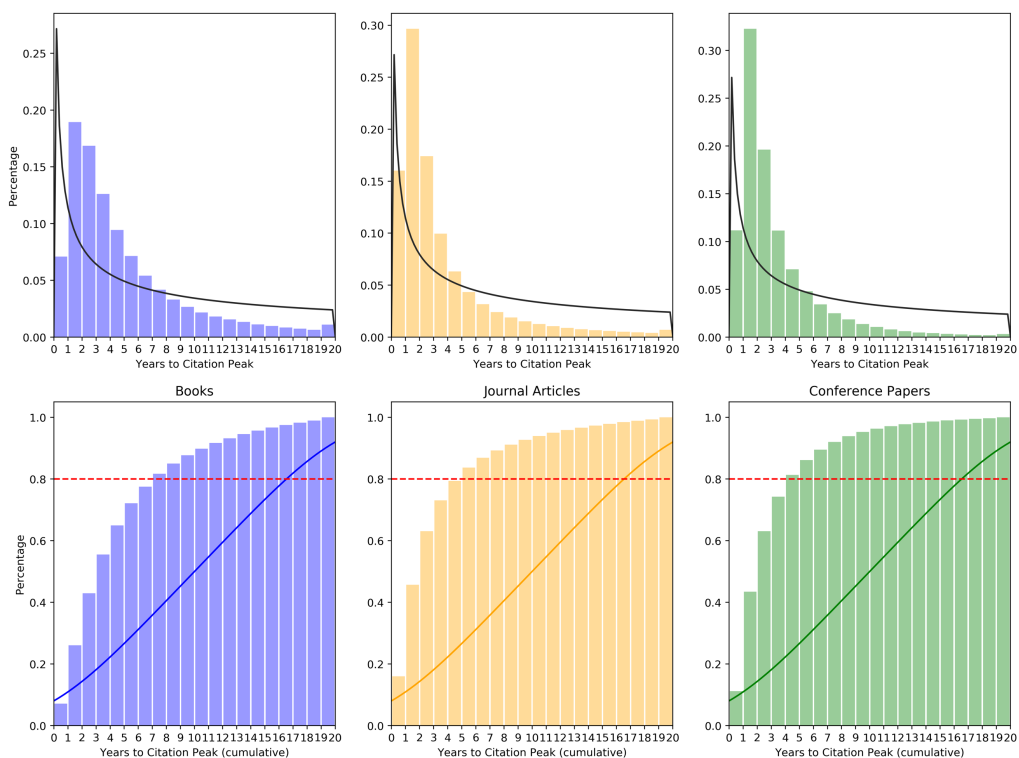

Figure 4. Years to Citation Peak (frequencies and cumulative frequencies)

Overall, books took longer to reach their citation peak than journal articles and conference papers. Some 80% of books took up to eight years to reach their citation peaks. On the other hand, it took only five years for 80% of journal articles and conference papers to reach their citation peaks. One reason could be that books have longer publication cycles than journal articles and conference papers. Frequency distributions of the three publication types were power law with different intensities. Overall, books had a gentle slope, whereas journal articles and conference papers had steeper slopes.

While years to citation peak measures how quickly or slowly citations to a book peak and drop off, cited half-life measures how long a book has received citations. Figure 5 shows frequency and cumulative frequency distributions of cited half-life, from one to twenty.

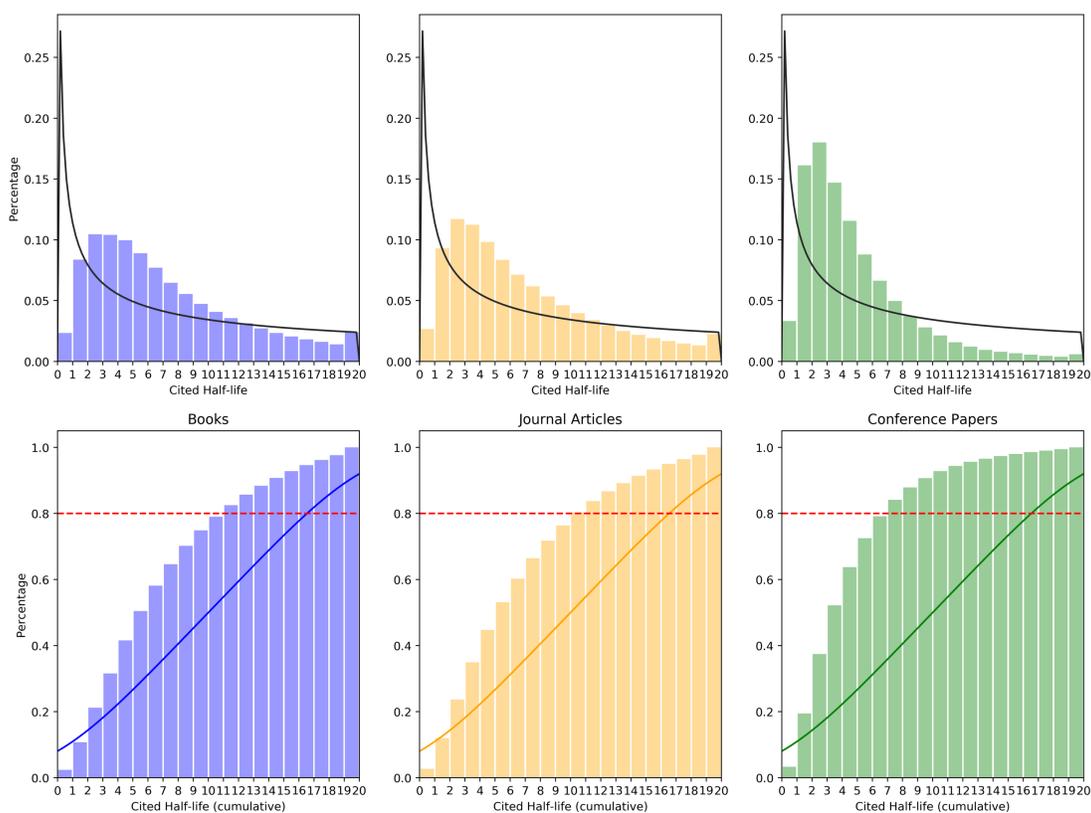

Figure 5. Cited Half-Life (frequencies and cumulative frequencies)

In Figure 5, we can see that books and journal articles had similar patterns, both in terms of frequency and cumulative frequency distributions. In frequency distributions, they followed a long-tailed distribution pattern, with gradual slopes. In cumulative frequency distributions, if we consider the percentage up to the 80% level, they also had a similar level of cited half-life, i.e., 11 years. These patterns differed from those of conference papers. This can be perceived as an indication that books and journal articles usually receive citations for a more extended period than conference papers. However, there was not a clear difference between books and journal articles.

*Disciplinary Differences in Book Citation*

Books of different disciplines may exhibit different citation characteristics. To understand disciplinary differences in book citation, we classified the books into multiple disciplines. We used Online Computer Library Center (OCLC)'s Classify service (http://classify.oclc.org/classify2/) to associate each book to one of the 21 subject areas of Library of Congress Classification (LCC) by matching ISBNs. We further grouped such 21 subject areas into five Web of Science (WoS) areas: *Arts & Humanities*, *Life Sciences & Biomedicine*, *Physical Sciences*, *Social Sciences*, and *Technology*. We considered only the Crossref types "book" and "monograph" in the analysis. Figure 6 shows differences among disciplines and book types.

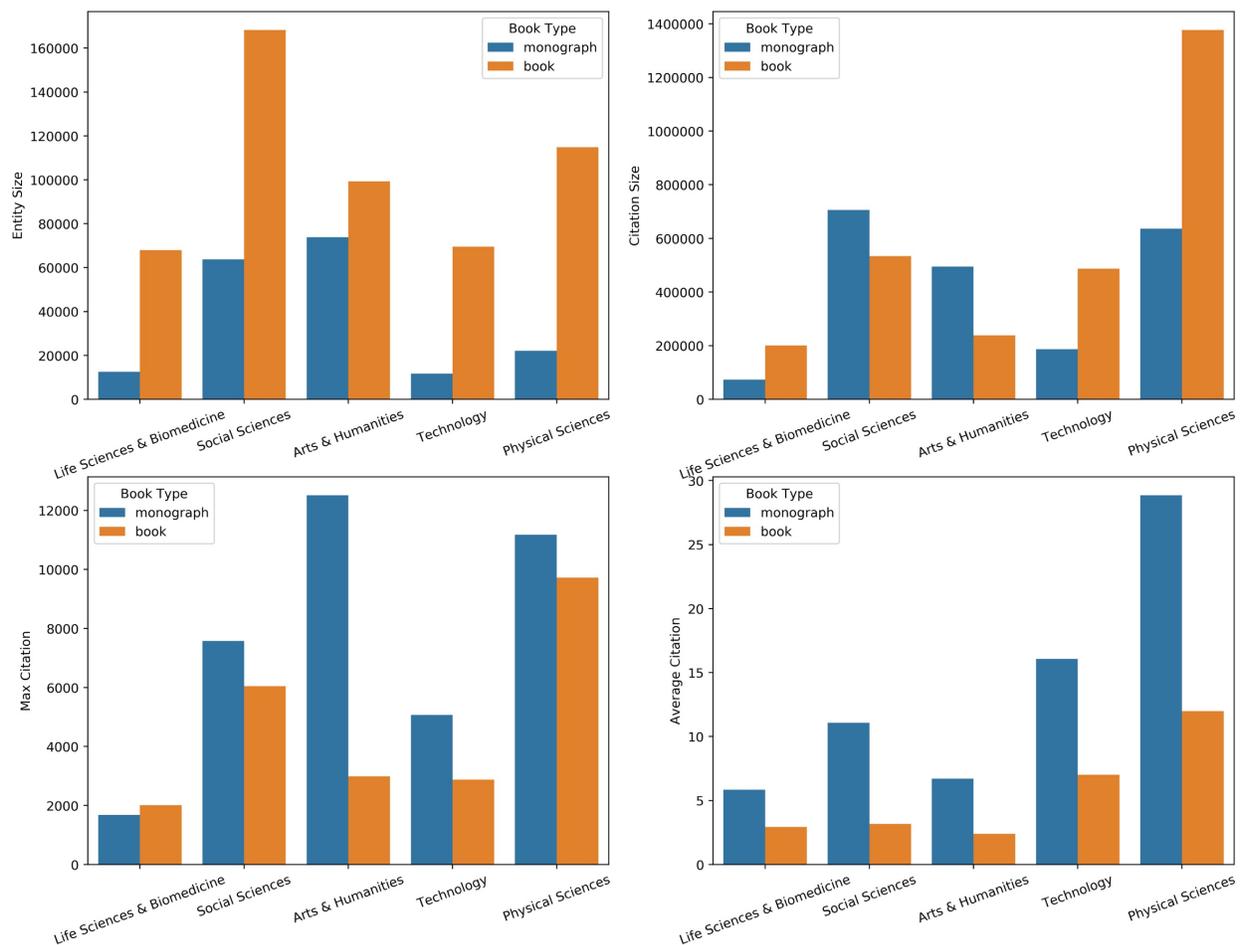

Figure 6. Citation Characteristics by Different Disciplines and Book Types

We found that books and monographs had different citation characteristics as reported in the earlier studies (Leydesdorff & Felt, 2012b; Gorraiz, Purnell, & Glänzel, 2013).

In the five WoS areas, books were published much more than monographs. In terms of the total citations each book type received, monographs received more citations than books in Social Sciences and Arts & Humanities. The salient point in our analysis is that, in Social Sciences, the size of book publication was double of that of monographs, though received fewer citations. In terms of the max and average citation, monographs received more citations than books with an exception in Life Sciences & Biomedicine, where books recorded a slightly higher maximum citation count. In our data, Social Sciences published the more significant number of books though received relatively fewer citations as we can see from the measure of average citation. Physical Sciences is the area that recorded the highest average citation in both books and monographs. In Arts & Humanities, there might be a considerable gap between the highly and regularly cited monographs as we noticed from the low average citation value. There might be a few influential monographs in the area. The disciplinary difference is also shown in average years to citation peak and cited half-life shown as in Figure 7.

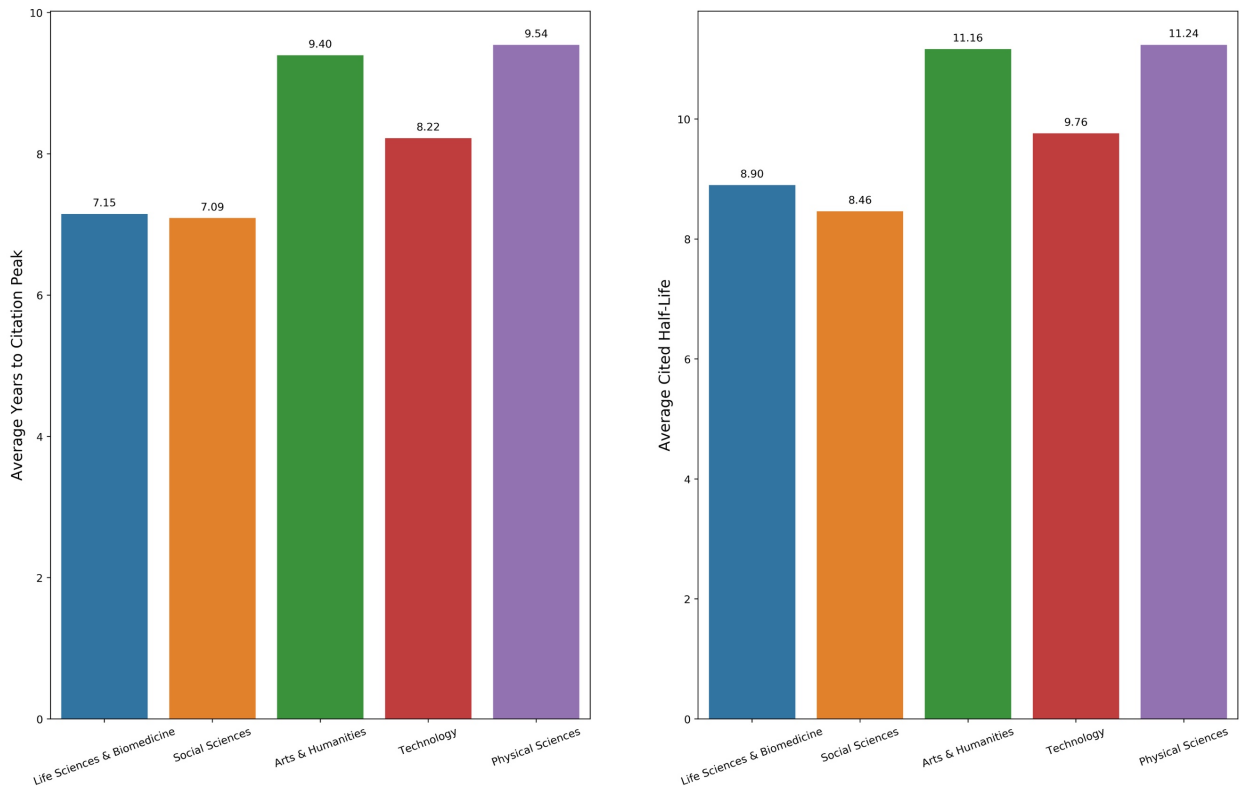

Figure 7. Average Years to Citation Peak and Cited Half-Life of the Five WoS Areas

In the five areas, Arts & Humanities and Physical Sciences had similar patterns. It took longer in the two areas to reach citation peak and cited half-life. Technology takes approximately one year shorter than the two areas mentioned above. Life Sciences & Biomedicine and Social Sciences showed highly similar patterns. These two areas took the shortest years to reach citation peak and cited half-life. We can see that citations to books of Life Sciences & Biomedicine and Social Sciences peaked more quickly while books of Arts & Humanities and Physical Sciences received citations for more extended periods than the other areas. Technology was positioned somewhere between the two groups. These results are in line with previous findings on different citation characteristics in different fields (e.g., Glänzel & Schoepflin,1999).

*The Twenty Most-Cited Books*

Table 4 shows publication year, title, and citations of the twenty most-cited books in the dataset.

Table 4. The Twenty Most Cited Books

| Publication Year | Book Title | Total Citations | Citations from Books | Citations from Conference Papers | Citations from Journal Articles |
|---|---|---|---|---|---|
| 1991 | Situated Learning | 12,513 | 11.96% | 3.86% | 75.63% |
| 1991 | Elements of Information Theory | 11,172 | 3.33% | 34.65% | 50.42% |
| 1995 | The Nature of Statistical Learning Theory | 9,724 | 6.11% | 41.82% | 42.31% |
| 1998 | Communities of Practice | 9,226 | 12.67% | 3.78% | 70.84% |
| 1994 | Linear Matrix Inequalities in System and Control Theory | 8,853 | 1.56% | 39.74% | 47.57% |
| 1965 | Society and the Adolescent Self-Image | 8,605 | 1.36% | 0.15% | 92.71% |
| 1997 | Processing of X-Ray Diffraction Data Collected in Oscillation Mode | 8,516 | 0.31% | 0.04% | 99.58% |
| 1993 | An Introduction to the Bootstrap | 8,317 | 3.82% | 4.35% | 86.32% |
| 1989 | Generalized Linear Models | 7,665 | 4.03% | 2.37% | 87.70% |
| 1977 | Outline of a Theory of Practice | 7,599 | 12.24% | 0.26% | 85.00% |
| 1990 | Institutions, Institutional Change and Economic Performance | 7,576 | 13.37% | 0.92% | 82.23% |
| 1992 | Ten Lectures on Wavelets | 7,102 | 3.87% | 21.06% | 59.39% |
| 1985 | Matrix Analysis | 6,132 | 1.99% | 24.93% | 65.15% |
| 1989 | Structural Equations with Latent Variables | 6,048 | 2.33% | 1.92% | 93.42% |
| 1981 | Pattern Recognition with Fuzzy Objective Function Algorithms | 5,680 | 7.11% | 38.40% | 39.07% |
| 1986 | Density Estimation for Statistics and Data Analysis | 5,258 | 5.00% | 12.29% | 74.06% |
| 2005 | Fundamentals of Wireless Communication | 5,069 | 2.60% | 50.42% | 46.14% |
| 1990 | Governing the Commons | 5,065 | 12.87% | 1.50% | 82.86% |
| 1958 | The Psychology of Interpersonal Relations | 4,929 | 4.61% | 0.83% | 92.45% |
| 1990 | Amplification and Direct Sequencing of Fungal Ribosomal RNA Genes for Phylogenetics | 4,927 | 0.22% | 0.26% | 96.83% |

Among these twenty books, the most recent (*Fundamentals of Wireless Communication)* was published in 2005, and the oldest (*The Psychology of Interpersonal Relations)* was published in 1958. From the list, one book was published in the 2000s, eleven books were published in the 1990s, and five books were published in the 1980s. The remaining books were published between 1950 and 1980. From the books' titles, we can summarize their three characteristics as follows:

First, they covered general topics, which were not very specific even if essential. Instead, they had broad applications in many fields. Second, they covered theories and technological concepts. Specifically, they mostly concerned theories, techniques, and algorithms. Finally, their major topics were mathematics and statistics. Considering the importance of these two subjects to modern science, this was an expectable result.

Most citations were from journal articles and conference papers, with only five books receiving more than 10% of their total citations from other books. We found that none of these five books (*Situated Learning; Communities of Practice; Outline of a Theory of Practice; Institutions, Institutional Change and Economic Performance;* and *Governing the Commons*) covered technological topics. While the highest proportion of citations to books came from journal articles, we found a few books – i.e., *Elements of Information Theory, The Nature of Statistical Learning Theory, Linear Matrix Inequalities in System and Control Theory, Pattern Recognition with Fuzzy Objective Function Algorithms,* and *Fundamentals of Wireless Communication* – which were highly-preferred by conference papers as cited material. More than 30% of their citations were from conference papers. All of the above books covered theories and algorithms, which is compliant with the citation behavior in conferences we identified since they usually seemed to focus on technological development.

**Discussion and Conclusions**

In this study, we investigated the roles of books in scholarly communication using the citation data included in COCI, created by OpenCitations. Using this dataset allowed us to present a clear and comprehensive overview of the roles that books play in scholarly communication. Our findings can be summarized as follows. Of the three major bibliographic types (i.e. book items, journal articles, and conference papers), books have received less than 4% of total citations. Before the 2000s, books received most of their citations from journal articles. Recently, however, this percentage has dropped since books received more citations by conference papers and other books. Besides, among the investigated books, 97.96% have been cited fewer than ten times. In contrast, 1.69% of books have received from ten to fifty citations.

We did not find a clear pattern regarding the relationship between books' publication years and their average citation count. Conversely, in the other two bibliographic types (i.e. journal articles and conference papers), entities published recently received more citations than older ones. Further exploration revealed that books took longer to reach citation peak than journal articles and conference papers. In paritcular, 80% of books took up to eight years to reach their citation peaks, while 80% of journal articles and conference papers reached their peak within five years. The analysis of the twenty most cited books revealed that they covered general (yet essential) topics, theories, and technological concepts in the fields of mathematics and statistics.

A comprehensive citation dataset (including books and other types of research publications) is a crucial tool for a correct understanding of books' place in scholarly communication. By using an open citation data set, this study demonstrated how to use open science datasets to further scientific research. The results not only revealed books' roles in scholarly communication but also showcased citation analysis methods that are independent of proprietary data sources. OpenCitations is a leading scholarly infrastructure organization enabling researchers to access and use open, high-quality citation data freely. Moreover, it represents a paradigm shift towards open, transparent, and reproducible research. Its goals echo those of sister initiatives, such as Wikidata (https://www.wikidata.org), and the Initiative for Open Citations (https://i4oc.org).

Finally, the findings shown in the study are representative, as COCI is one of the largest open citation data repositories. Books are a more traditional form of writing than journal articles and conference papers. Given that the latter two have been used mainly in scientific writing, as we have shown in the study, books present different characteristics. Overall, books have limited (but irreplaceable) roles to play in scholarly communication. Books cover essential theories and technologies that are fundamentals for scientific research in greater detail and with compressive coverage. Such vital information cannot be conveyed in other formats, e.g., journal articles. In scholarly communication, books are an indispensable scientific asset which continually make valuable contributions.

## Acknowledgements

This paper was supported by Sungkyun Research Fund (S-2018-2538-000), Sungkyunkwan University, 2018.